\documentstyle[12pt]{article}

\begin{document}

\begin{center}

{\Large\bf Enhancing Efficiency of Mixing in Chaotic Flows}

\end{center}

\vspace{10pt}

\begin{center}
Neelima Gupte and R.E. Amritkar

Department of Physics, University of Poona

Pune---411007, INDIA.

\end{center}

\vspace{10pt}

\begin{abstract}
We propose a mechanism by which the efficiency of mixing  in chaotic flows
can be enhanced. Our mechanism consists of
introducing small changes in the system parameters
in regions of phase space where the local Lyapunov exponent falls 
substantially below its average value. We have applied our mechanism 
to several typical 
chaotic maps and flows including a system of chemical reactions. We find that
our method is quite efficient as it gives a substantial enhancement of 
the rate of mixing with
small changes in system parameters, without disturbing the attractor 
significantly.

\end{abstract}
\vspace{10pt}
PACS No. 05.45.+b
\newpage

The importance of mixing has long been recognised in the context of physical
systems \cite{O}. Examples of mixing processes can be found in the context of 
combustion processes \cite{WWH}, fluid flows \cite{E,W,Z}, viscous liquids
\cite{SW}, chemical reactions \cite{WR,AC}, heat transfer processes
\cite{T} etc. Again, efficient mixing has desirable consequences in many
practical contexts. A well mixed fuel--air mixture can lead to greater 
efficiency of the combustion process. Similarly, if the  reactants 
of a chemical reaction are mixed better, this can lead to better yields of
the resultants. Heat
transport in convective processes can be enhanced by improved mixing.
Many of these mixing processes 
can be modelled by chaotic flows \cite{O}. Hence, if a mechanism can be 
found for the enhancement of the rate of 
mixing in chaotic flows, it  can prove to be 
very 
useful in a variety of contexts. We propose such a mechanism in this paper.

Mixing is a consequence of the stretching and folding of chaotic flows.
A system which has exponential stretching, as in a chaotic flow,
mixes efficiently \cite{O}. For a chaotic flow, the average rate of stretching 
can be characterized by the Lyapunov exponent. However, the rate of 
stretching is 
not uniform over the chaotic attractor.
Thus the 
local Lyapunov exponent 
(LLE), a measure of the local rate of stretching,  
is different in different regions of the attractor \cite{ABAR}. 
We exploit the nonuniform
nature of the spatial distribution of the local Lyapunov exponents to
construct a mechanism that can enhance the rate of 
chaotic mixing. Briefly, we enhance 
the average rate of stretching by introducing a small parameter 
perturbation which enhances the local Lyapunov exponent whenever the system 
trajectory visits a region where the local Lyapunov exponents take
values much smaller than their average value. We find that this procedure 
works quite efficiently as small perturbations in parameter made for small 
times compared to the total time of evolution can lead to substantial
enhancement of the Lyapunov exponent and thus the mixing efficiency.

Let us consider an autonomous nonlinear dynamical system of dimension $n$,
evolving via the equations
$\dot{\bf x} = {\bf F}({\bf x}, \mu)$,
where the set of parameters $\mu$ takes  values such that the trajectory 
shows 
chaotic behaviour. Let ${\bf w}({\bf x}, t)$ be the tangent vector to the
trajectory at the point ${\bf x}$ and time $t$. The evolution of ${\bf w}$ is 
given by 
$\dot{\bf w} = ( {\bf x \circ \nabla}) {\bf F}$. 
The Lyapunov exponent of the system is defined by
\begin{equation}
\lambda = \lim_{t\to \infty} {1 \over t} \ln {||{\bf w}({\bf x},t)||
	     \over ||{\bf w}({\bf x}(0), 0)||}.
\label{lambda}
\end{equation}
where ${\bf x}(0)$ is the value of ${\bf x}$ at $t=0$. We now define the local 
Lyapunov exponent $\lambda({\bf x})$ 
as
\begin{equation}
\lambda({\bf x}) = \lim_{\Delta t \to 0} {1 \over \Delta t} \ln {||
{\bf w}({\bf x}(t+\Delta t), t+\Delta t)|| \over ||{\bf w}({\bf x}(t),t)||}
\label{lle}
\end{equation}
Clearly $\lambda({\bf x})$ represents the local rate of stretching at the
point ${\bf x}$. This is, in general, not uniform over the attractor.
We also note that the Lyapunov exponent 
$\lambda$ (Eq.(\ref{lambda})) is the average  value of
the local Lyapunov exponents for a long orbit.

We set up a control procedure to enhance the mixing efficiency 
utilising the distribution
of the local Lyapunov exponents. The control procedure operates in regions 
where the LLE-s fall substantially below the average value $\lambda$. 
If, at any time,
the local Lyapunov exponent of the system falls below its average value 
to the point where
\begin{equation}
\lambda({\bf x}) < (\lambda - \gamma \sigma_\lambda) 
\label{condition}
\end{equation}
where $\sigma_\lambda$ is the
standard deviation of the distribution of LLE and $\gamma$ is some chosen 
factor,
the control is activated so that 
the parameter $\mu$ is changed to $\mu + s \, d\mu$. Here $d\mu$ is a small 
increment and $s$ takes values $+1$ or $-1$ depending on which choice enhances
the LLE. The system is allowed to evolve with the new value of the parameter
as long as  the condition (\ref{condition}) is satisfied. Thereafter the  
parameter is reset to its original value.

To decide the sign $s$, we write equation for ${\bf w}$ 
in  matrix notation in the form
\begin{equation}
\dot{W}^T = W^T M^T, \; \; \dot{W} = M W ;
\label{matw}
\end{equation}
where $W^T$ is a row vector and the matrix $M^T$ is given by
$ M^T = { \bf \nabla \; F}$.
The equation for the norm of $ W $ can be written as
\begin{eqnarray}
\dot{||W||^2} & = & W^T ( M^T + M ) W
\end{eqnarray}
Thus the rate of change in the norm of $ W $ due to change in the 
parameter is given by
\begin{eqnarray}
\Delta \dot{||W||^2} & = & \dot{||W(\mu+d\mu)||^2} - \dot{||W(\mu)||^2} 
\nonumber \\
& \simeq & W^T ( M^T_\mu + M_\mu ) W \, d\mu
\label{deltaw}
\end{eqnarray}
where the last step is obtained by expanding to lowest order in $d\mu$ and
$ M_\mu = \partial M / \partial \mu$. Clearly, for the local rate of 
stretching to increase, $\Delta ||W||^2$ must be positive . 
Thus the sign $s$ is determined to ensure
that $\Delta ||W||^2$ is positive. 

It must be noted that Eq.(\ref{deltaw}) is written in the lowest 
order in $d\mu$.
Actually, the effect of the perturbation is nonlinear since 
 when the parameter changes the entire trajectory of the system changes. 
Hence the effect on the LLE can be 
quite different from that given by Eq.(\ref{deltaw}) 
due to the effect of the higher nonlinear 
terms. In many cases the 
enhancement in the Lyapunov exponent turns out to 
be substantially higher than 
that expected
in the linear approximation.

We now illustrate our procedure using some typical flows. We first consider 
the Lorenz system \cite{L} with parameters $\sigma = 10.0, \; r=30.0$ and $b =
8/3$.
We choose $r$ as the control parameter. The perturbation is switched
on when the condition (\ref{condition}) is satisfied. The sign of the 
perturbation $dr$ is 
obtained using Eq.(\ref{deltaw}) and the sign is decided by
$ W_xW_ydr > 0$. For a change of parameter $dr=1.0$, $\gamma=0.5$, the Lyapunov 
exponent of the system 
is enhanced from $\lambda=0.950$ for the uncontrolled case to $\lambda=
1.440$ (See Table I).

This enhancement in the Lyapunov exponent is 
not confined to the parameter  values above. The plot of the  Lyapunov
exponent of the system
as a function of $r$ for both the uncontrolled and the controlled case is 
shown in Fig. 1.
It is clear from the figure that there is a substantial  enhancement of the 
Lyapunov exponent over the entire range plotted in the figure .
This enhancement has been effected by causing a change in the local Lyapunov
exponents of the system via parameter change.
To show this, we plot
the distribution of the LLE for $r=30.0$ for both the uncontrolled and the 
controlled cases in Fig.2. It is clear that the distribution of local 
Lyapunov 
exponents of the system
has changed in a manner in which the average exponent is significantly
enhanced.

In order to show that the increase in the Lyapunov exponent 
translates into an enhancement of mixing efficiency, we operated 
the control procedure on a large number of initial conditions
in a small region of phase space. We cover the attractor with 
a grid of cubic boxes. One of the box is chosen randomly. We take a
large number of initial conditions in this box. Each initial condition 
is evolved according to our control algorithm i.e. the control is operative
for a given trajectory (corresponding to a given initial condition)
whenever the local Lyapunov exponent of the trajectory 
satisfies condition (3). The initial conditions are also evolved 
separately without the control. The initial conditions initially in 
one box spread over several boxes with time. A comparison of the number
of occupied boxes, i.e. the boxes which have at least one initial
condition, as a function of time for the uncontrolled and the
controlled systems gives us an idea of the relative rates of mixing 
of the two systems. Fig.3 plots the number of occupied boxes
as a function of time for both the uncontrolled and the controlled
systems with a grid of $10^3$ boxes and $10^5$ initial conditions.The 
parameter values are $\sigma=10.0, r=30.0, b=2.6666, dr=1.0$ and $\gamma=0.25$
\cite{FN}.
It is clear from the figure that the controlled system mixes at a
faster rate than the uncontrolled one. The results are unchanged 
for any randomly chosen initial box. This  demonstrates that 
the control procedure has successfully enhanced the mixing efficiency of 
the system.

In order to demonstrate the efficacy of our procedure for a system 
of the type that constitutes our motivation, we apply our mixing
algorithm to the Williamowski-Rossler attractor,  which models 
a system  of chemical reactions \cite{WR,AC}. The Williamowski-Rossler system 
evolves via the system of equations   
\begin{eqnarray}
\dot{x} &=& k_1x-k_{-1}x^2 -k_2xy+k_{-2}y^2 -k_4xz+k_{-4}\nonumber \\
\dot{y} &=& k_2xy-k_{-2}y^2 -k_3y+k_{-3} \nonumber\\
\dot{z} &=&-k_4xz+k_{-4}+k_5z-k_{-5}z^2
\label{Willross}
\end{eqnarray}
The system is allowed to evolve at the parameter values $ k_1=30.0, k_2=1.0,
k_3=10.0,k_4=1.0,
 k_5=16.5,
k_{-1}=0.25,
k_{-2}=1.0 \times 10^{-4},
k_{-3} = 1.0 \times 10^{-3}=k_{-4}, k_{-5}=0.5$  . Control is effected via a
 change in parameter $k_1$ whenever the 
condition (\ref{condition}) is satisfied. As seen from Table I this results
in a large enhancement of the Lyapunov exponent from the uncontrolled value
$\lambda=0.559$ to the value $\lambda=0.804$ after the application 
of the control. We have also verified that the rate of mixing is enhanced 
due to the control by evolving a large number of initial conditions 
in a small region of phase space.

The mixing procedure discussed above can be easily modified to apply to the 
case of maps. In this case the condition (\ref{deltaw}) gets modified to
the form 
\begin{eqnarray}
\Delta {||W||^2} & = & W^T (M M^T_\mu + M_\mu M^T ) W \, d\mu
\label{deltaw1}
\end{eqnarray}
For control to enhance mixing efficiency, 
the parameter change $d\mu$ must be such that
$\Delta{||W||^2}$ is positive. If this procedure is applied to the Henon map
\cite {Hen} at parameter values $a=1.2, b=0.3$ with $da=0.1$ as the parameter
change, we again find an enhancement of the Lyapunov exponent from the
uncontrolled value $\lambda=0.306$ to the controlled value $\lambda=
0.328$ (See Table I).

The increase in the rate of spread of initial conditions in phase space
of the controlled Henon system as compared to the uncontrolled one 
is demonstrated in Fig.~4  
 ($a=1.1, b=0.3, \gamma=0.5$ and
$da=0.2$).Figs 4(a) and 4(b) show the uncontrolled and controlled systems respectively after 10 iterations. At this stage itself the controlled system has spread out more than the uncontrolled situations. This difference can be even more clearly seen in Figs 4(c) and 4(d) which show the uncontrolled and controlled systems after 1500 iterations. The Lyapunov exponent has changed from $\lambda=
0.177$(uncontrolled) to $\lambda=0.258$ (controlled).

Thus our control procedure works for all the maps and flows tested including
the case of the chemical reaction system and the Lyapunov exponent is 
substantially enhanced in most cases. However, it is important to ensure
that the control does not disturb the attractor unduly. Visual comparisons 
of the appearance of the controlled and uncontrolled attractors reveal no
significant differences between  them. A more quantitative comparison can
be made by comparing the fractal (box-counting) dimensions of the two.
In the case of the Lorenz system, for the parameter values listed in Table
I, the fractal dimension of the uncontrolled attractor was $D_0=2.052$ whereas
that of the controlled attractor was $D_0=2.056$. In the case of the Henon
attractor, again for the values of Table I, the 
fractal dimension changed
insignificantly from $D_0=1.206$ to $D_0=1.212$. For the Williamowski-Rossler
attractor, the fractal dimension remained practically unchanged at $2.07$. 
Thus the 
control procedure does not appear to disturb the attractor unduly at the
present values of parameter change.

The Lyapunov exponent referred to in the entire 
discussion above is the largest
Lyapunov exponent of the system. A possible reason for the insignificant 
increase 
in dimension seen despite the application of control  can be found in the 
values of the other Lyapunov
exponents of the system. For the Lorenz attractor, for the parameter values
$\sigma=10.0, r=30.0, b=2.6666, dr =1.0$ and $\gamma=0.5$, the complete set of Lyapunov exponents of 
the uncontrolled system is 
given by
$(0.950, 0.000, -14.617 )$, and that  of the controlled system is  
given by 
$(1.440, -0.420, -14.686 )$.
Thus the largest Lyapunov exponent of the system, which is a measure of the 
rate of stretching, has increased whereas the
other  Lyapunov exponents of the system, 
 have become more negative  signifying an increase in the rate of 
contraction. The controlled system no longer has a zero Lyapunov exponent as
the trajectory is no longer smooth. These  observations are also
 true of 
the other
systems studied. The control procedure tends to push the trajectory in the 
basin of attraction of the uncontrolled attractor, but the increased rate of 
contraction pushes it back to the original attractor. Both these factors 
work to our advantage.
As mentioned  earlier, the increase in the rate of 
stretching tends to mix the system better. The  increase in the rate
of contraction has the advantage that it tends to stabilise the attractor,
so that the attractor is not unduly disturbed by the perturbation. This is 
the origin of the insignificant change in the fractal dimension of the
controlled and uncontrolled attractors. However, for large changes in
parameter the difference between the fractal dimensions  of the controlled 
and uncontrolled attractor does increase. Again  the stability 
of the uncontrolled attractor plays an important role in this. Although
the Lorenz attractor remains stable for large changes in parameter, and does
not show a large increase in dimension, the difference increases substantially
for the Henon attractor.     

The control procedure outlined above leads to an enhanced 
rate of mixing for most
parameter settings. However, we did find a few cases where it did not work 
well, e.g . in the neighborhood of the parameter values $r=138.0 $ and 
$r=160.0 $ for
the Lorenz attractor. This happened for parameter values where there was 
a wide  periodic window nearby. In such cases the control tends to push
the trajectory in the neighbourhood of a periodic orbit. 
 As a consequence the trajectory appears to show intermittent behaviour and the Lyapunov
exponent
does not  increase and sometimes even decreases. This problem can be
easily taken care of by changing the magnitude of the parameter change
and/or the factor $\gamma$.

Our control procedure works quite efficiently as it produces a substantial
enhancement of the Lyapunov exponent for quite small changes in the
parameters. This is due to the fact that our control procedure 
works by switching between three types of chaotic
flows, those characteristic of parameter values $ \mu $, $\mu +d\mu $ and
$ \mu-d\mu $. This switching introduces an extra time dependence in the
problem and is the origin of the efficiency of the procedure.

To summarise, we have introduced an efficient mixing mechanism that
produces a substantial increase in the rate of 
mixing for small changes in parameters.
The chaotic attractor is not disturbed unduly. The success of the mechanism
has been demonstrated for several chaotic flows and maps. We hope that this
mechanism will prove to be useful in enhancing the rate of
mixing in a variety of
practical contexts.

We thank the Department of Science and Technology(India) for financial
assistance. We thank IUCAA (Pune, India) and IMSc (Madras, India) for the
use of their computing systems.

\newpage
\begin{center}
{\large\bf Table Caption}
\end{center}
We list the uncontrolled (Free) and controlled (Cont.) 
values of the Lyapunov exponent
and of the fractal dimensions for several maps and flows. The values
of the parameters of the systems analysed are listed in the text.
The column Fract. refers to the fraction of time for which the system
is controlled and $d\mu$ is the parameter change.
\vspace{15pt}
\begin{center}
\begin{tabular}{lccccccc}
\hline
System & $\gamma$ & $d\mu$ & Fract. & \multicolumn{2}{c}{Lyapunov exp.}&
 \multicolumn{2}{c}{Dimension} \\
\cline{5-8}
 & & & & Free & Cont. & Free & Cont. \\
\hline
 Lorenz  & 0.5 & $dr=1.0 $ & 0.344 & 0.951 &1.440&
2.052& 2.056 \\
 & 0.25 & $dr=1.0$ & 0.447 & 0.951 & 1.362 & 2.052 & 2.061\\
Williamowski-Rossler & 1.0 & $ dk_1=1.5$
& 0.047 & 0.559 & 0.804 & 2.069 & 2.068 \\
 Henon & 0.5 & $da=0.1$ & 0.263 & 0.306 & 0.328 & 1.206 & 1.212 \\
\hline
\end{tabular}
\end{center}

\newpage

\noindent {\large\bf Figure Caption}

\vspace{10pt}

\begin{itemize}

\item[Fig.1] The plot of the Lyapunov exponent of the Lorenz attractor
for the parameters $\sigma=10.0, b=2.6666$ and $r$ from  $r=28.0$ to $r=80.0$.
The lower curve corresponds to the Lyapunov exponent for the uncontrolled 
system,
the upper to the controlled system with $\gamma=0.5$ and $dr=1.0$

\item[Fig.2] A histogram of the distribution of the local Lyapunov exponents
of the uncontrolled and controlled Lorenz systems for the parameter values
$\sigma=10.0, r=30.0$ , $ b=2.6666$,$\gamma=0.5 $ and $dr=1.0$. The data has been binned into ten
boxes. Each box is divided into two parts. The bar occupying the
left half of the box shows the normalised frequency of occurance
of the corresponding LLE of the uncontrolled system and the bar occupying
the right half of the box (the bar with vertical lines)
shows the same quantity for the controlled system.

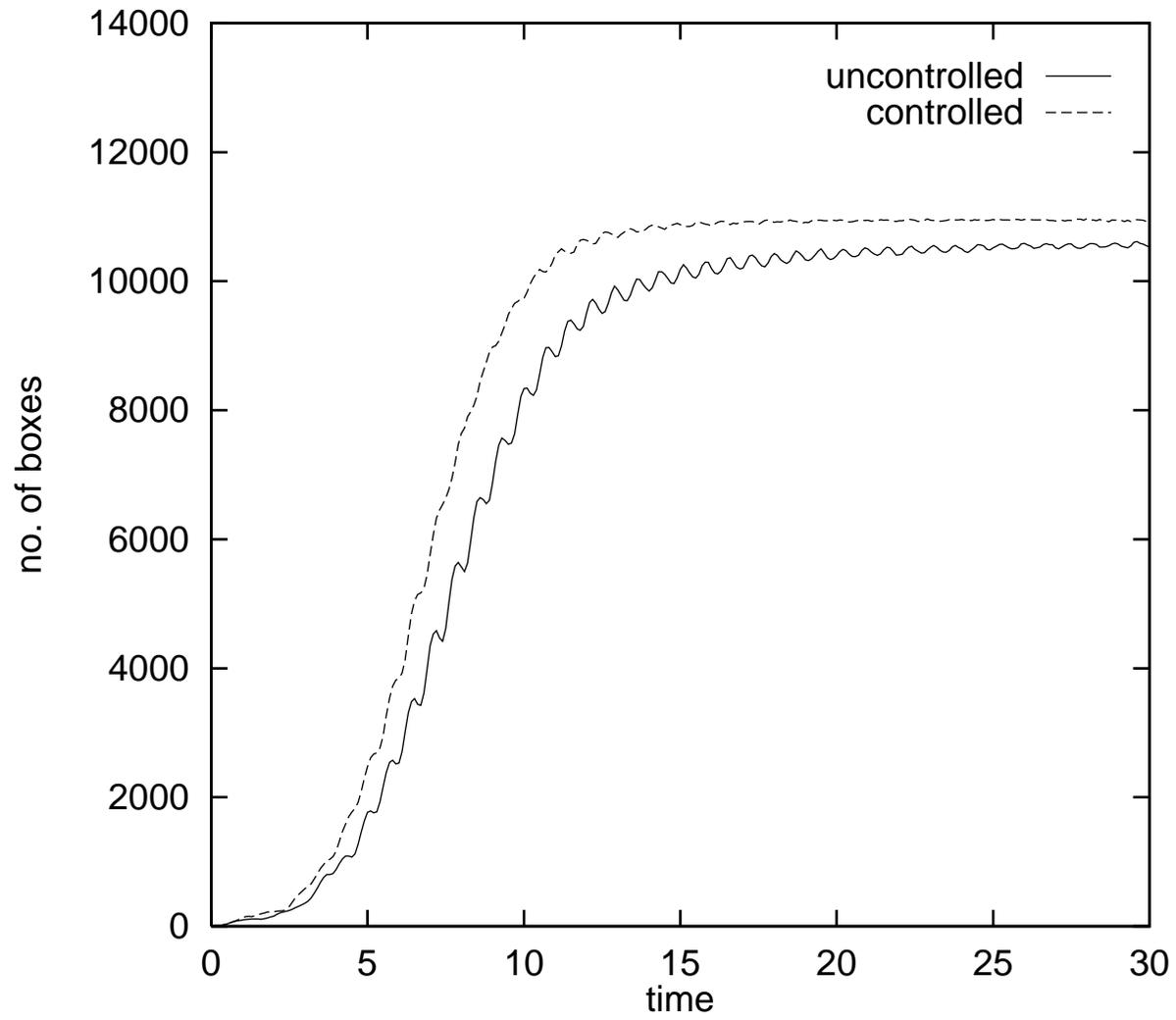
\item[Fig.3] The plot of the number of occupied boxes as a function of time
for the uncontrolled (solid line) and controlled (dashed line) Lorenz systems.
The parameter values are $\sigma=10.0, b=2.66666, r=30.0, \gamma=0.25$ and
$dr=1.0$.

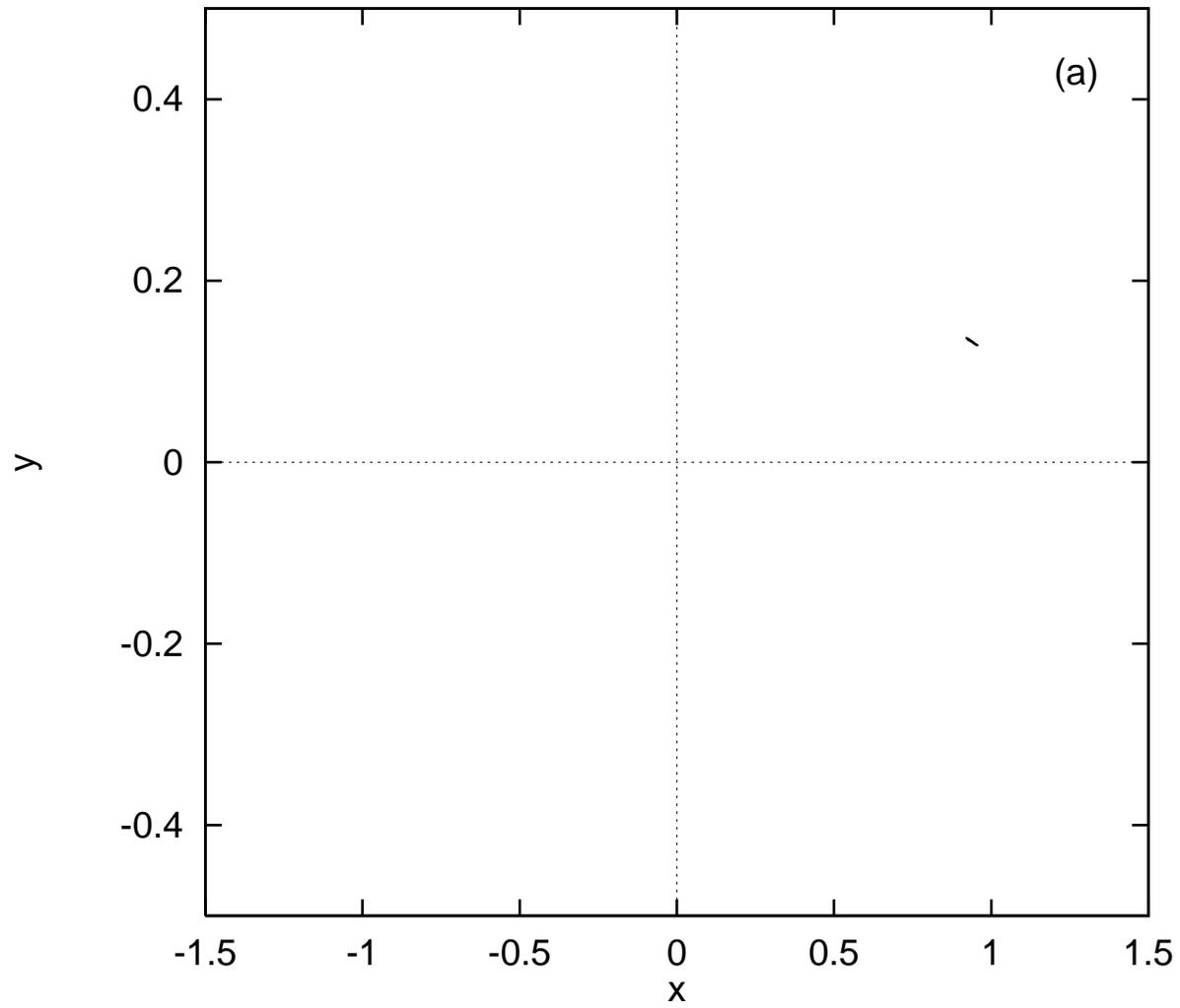
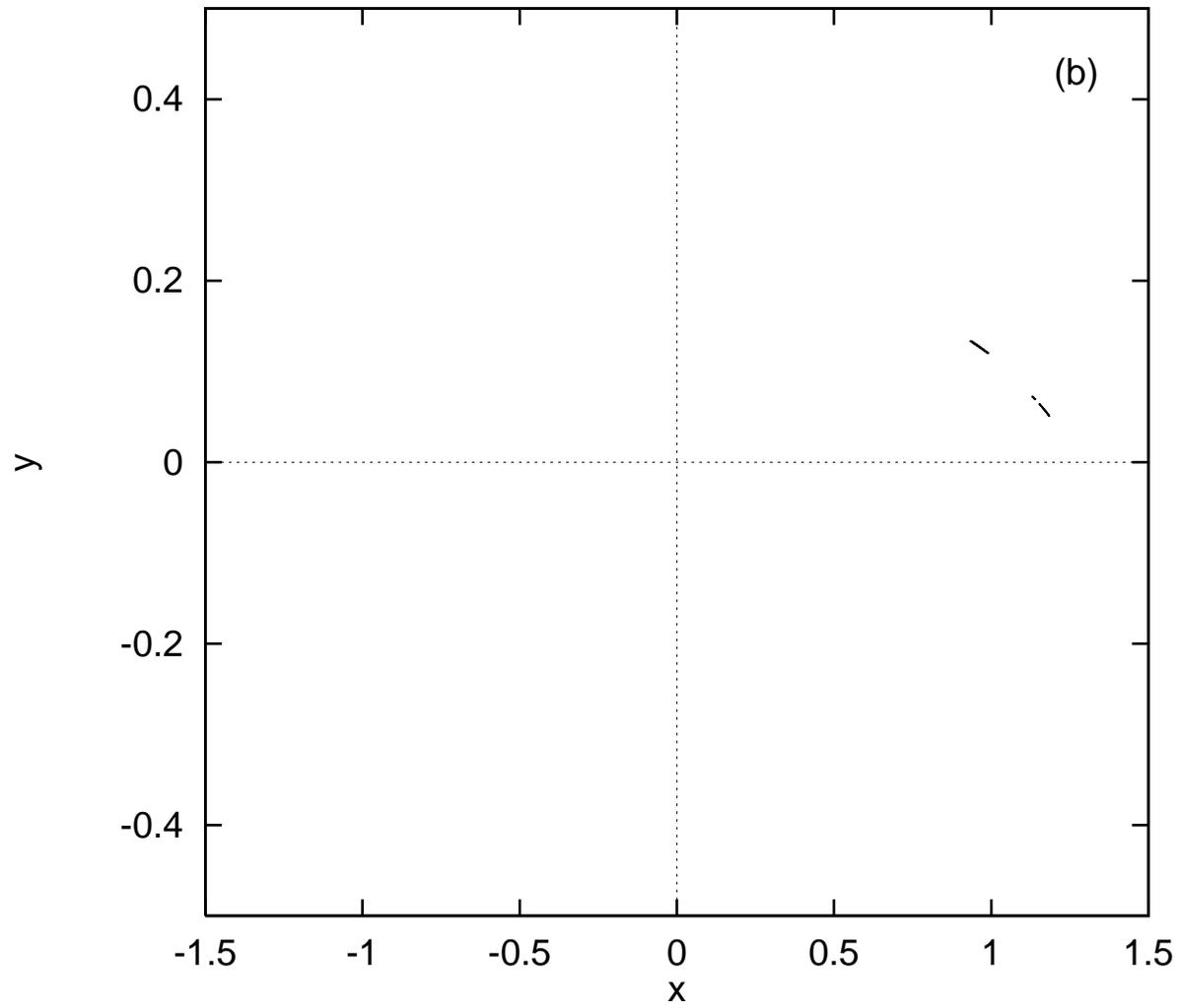
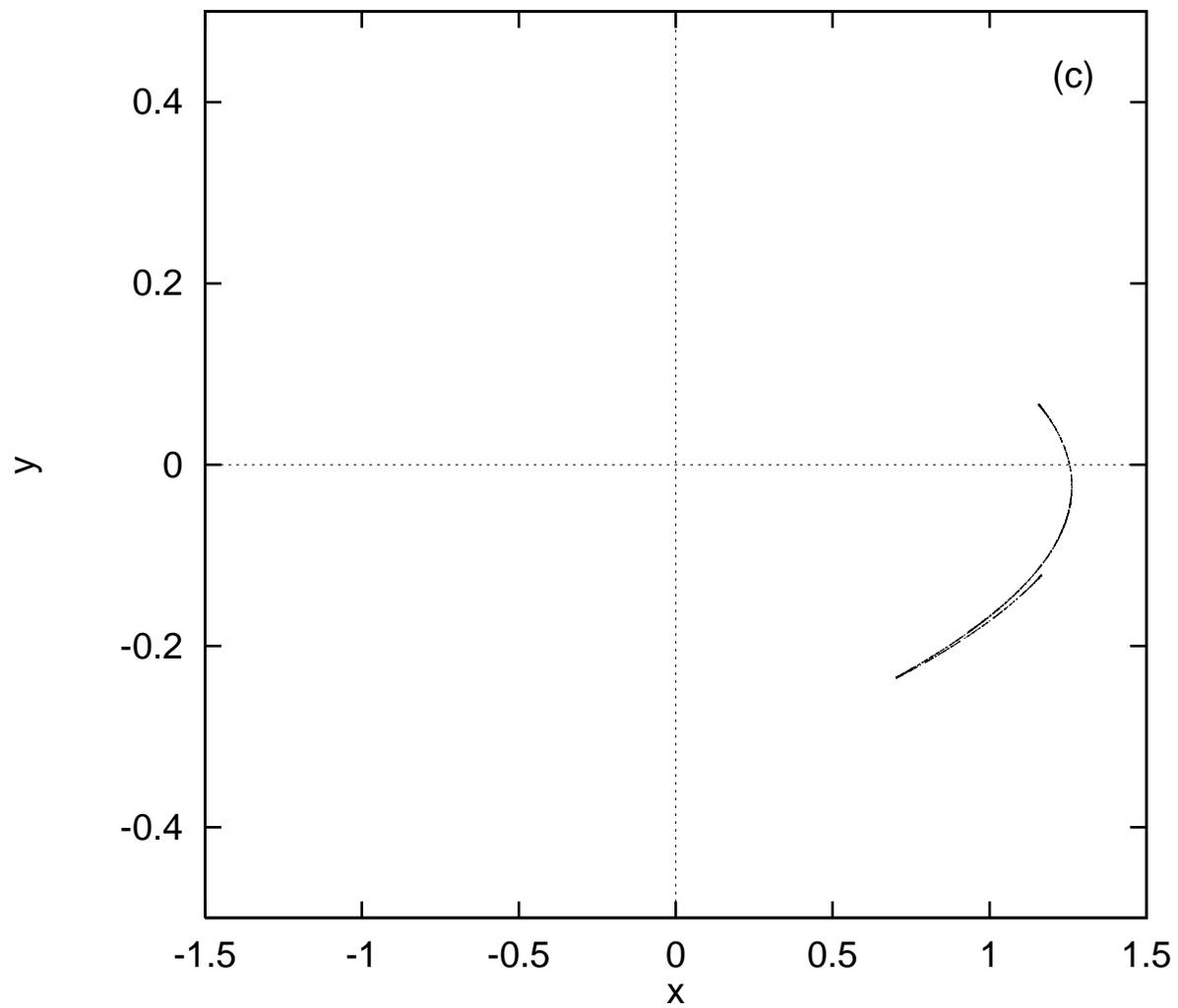
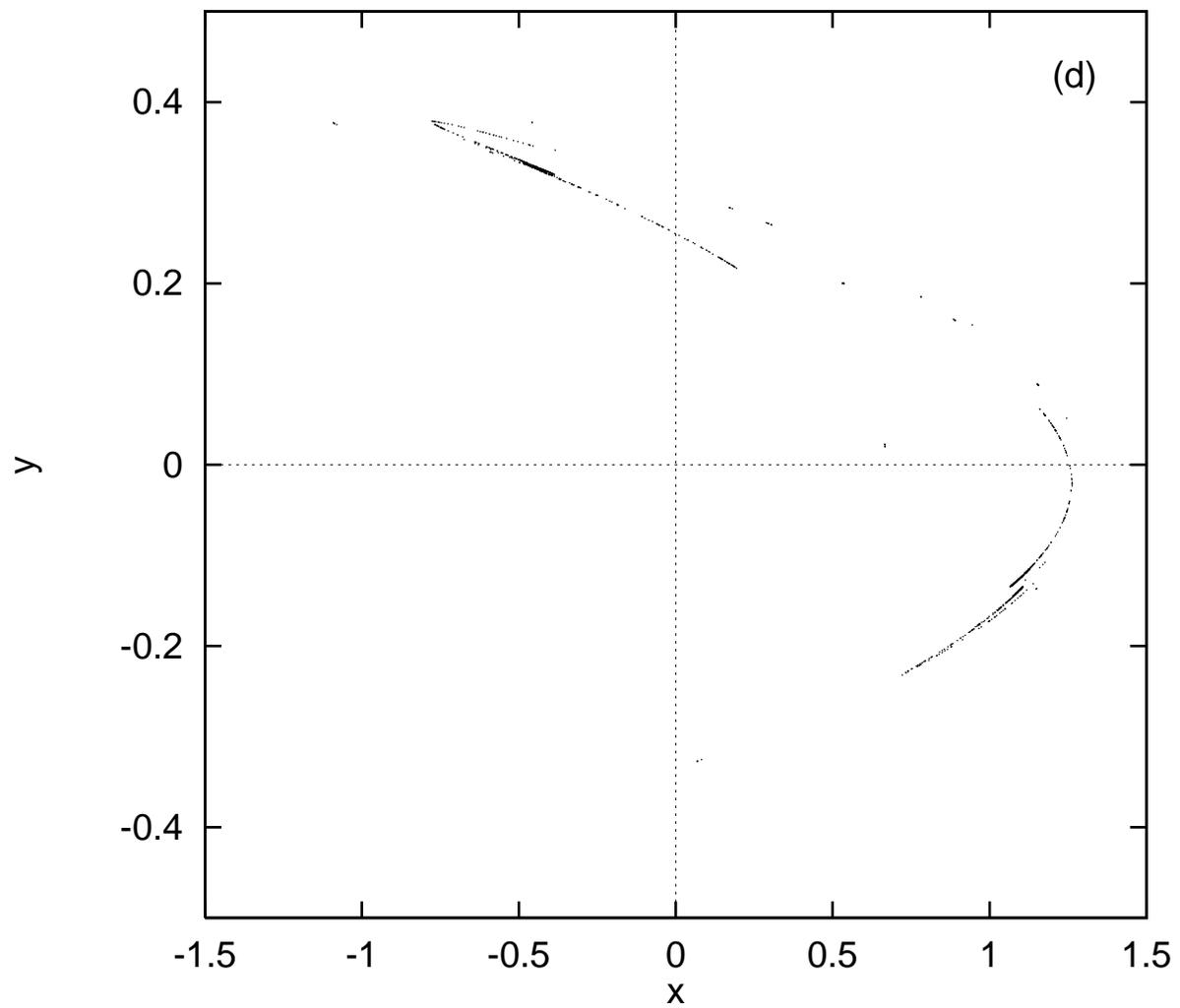
\item[Fig.4] We show the spread of 1000  points, initially in the same box
of a 128 by 128 grid on the Henon attractor. Fig 4(a) and Fig. 4(b) show
the uncontrolled and controlled Henon system after 10 iterates.
Fig. 4(c) and Fig. 4(d) show the uncontrolled and controlled Henon system after
1500 iterates. The parameter values are $a=1.1,b=0.3$ and $\gamma=0.5$
and $da=0.2$.

\end{itemize}


\begin{thebibliography}{99}
\bibitem{FN} For $\gamma=0.5$ it is necessary to take a much larger number  
of initial conditions to obtain a plot that is robust to change in initial conditions.
\end{thebibliography}
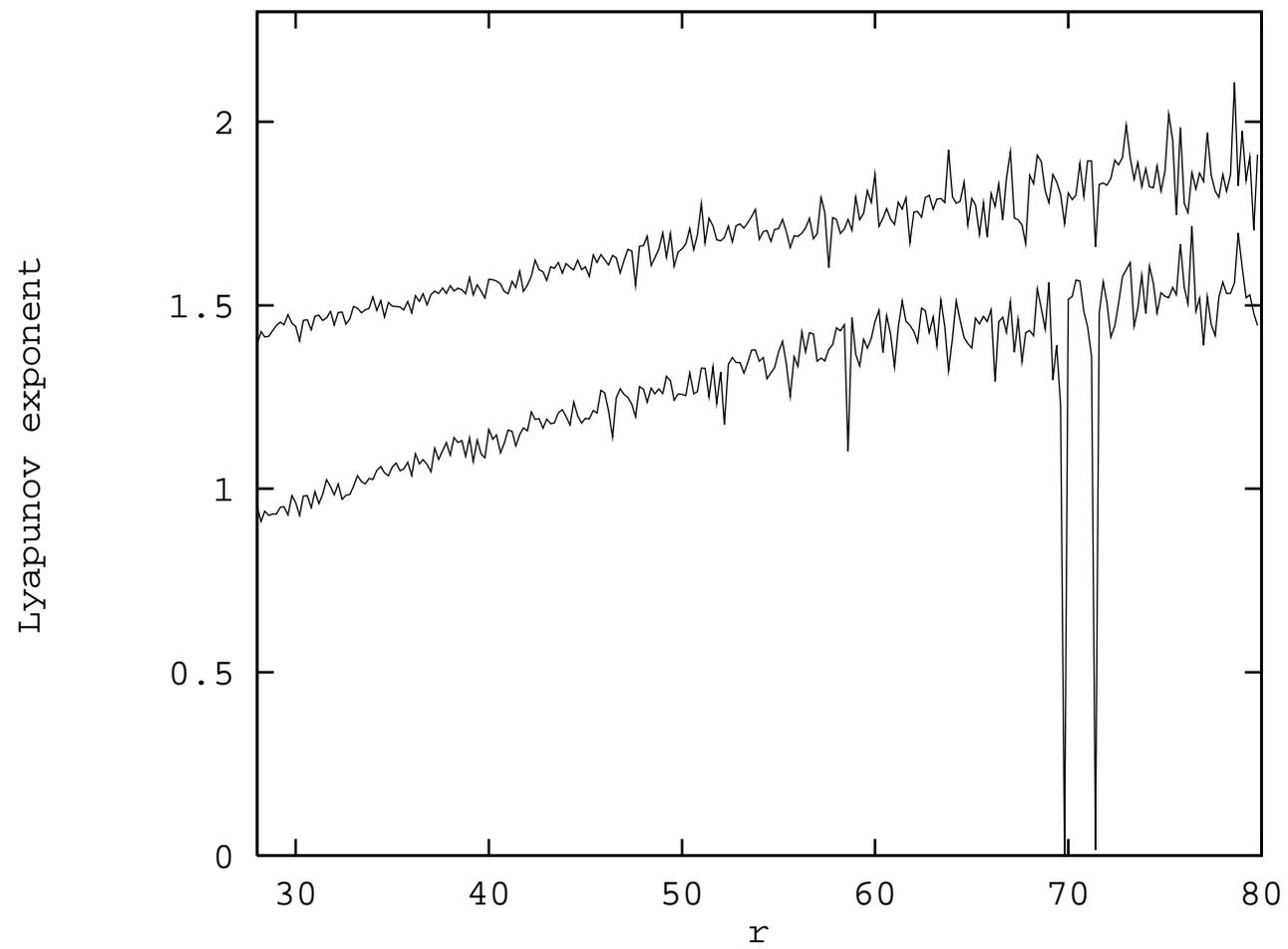
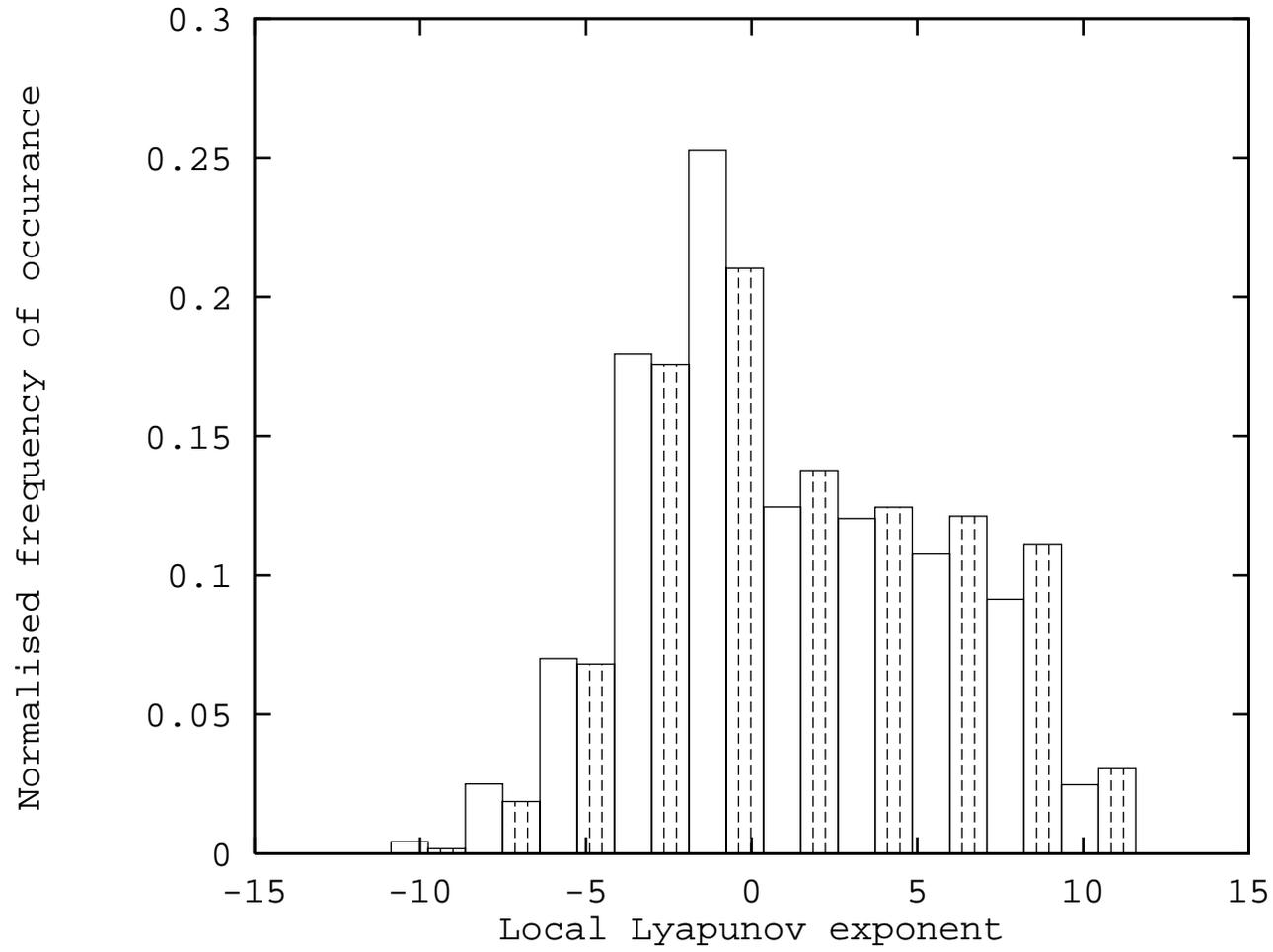
\end{document}